\documentclass[a4paper]{article}
\usepackage{textpos}
\usepackage{INTERSPEECH2020}
\usepackage{multirow}
\usepackage{subfigure}
\usepackage{enumitem}
\usepackage{svg}
\newlist{steps}{enumerate}{1}
\setlist[steps, 1]{label = Step \arabic*:}
\usepackage[ruled,vlined]{algorithm2e}
\usepackage[compress]{cite}
\usepackage{url}

\usepackage{enumerate}
\usepackage{subfigure}

\usepackage[compress]{cite}

\title{Improve Cross-lingual Voice Cloning Using Low-quality Code-switched Data}
\name{Haitong Zhang , Yue Lin }
\address{
  NetEase Games AI Lab, China
  }
\email{\{zhanghaitong01, gzlinyue\}@corp.netease.com}

\begin{document}

\maketitle

\begin{abstract}
Recently, sequence-to-sequence (seq-to-seq) models have been successfully applied in text-to-speech (TTS) to synthesize speech for single-language text. To synthesize speech for multiple languages usually requires multi-lingual speech from the target speaker. However, it is both laborious and expensive to collect high-quality multi-lingual TTS data for the target speakers. In this paper, we proposed to use low-quality code-switched found data from the non-target speakers to achieve cross-lingual voice cloning for the target speakers. Experiments show that our proposed method can generate high-quality code-switched speech in the target voices in terms of both naturalness and speaker consistency. More importantly, we find that our method can achieve a comparable result to the state-of-the-art (SOTA) performance in cross-lingual voice cloning. 

\end{abstract}

\section{Introduction}

Recently, text-to-speech (TTS) has witnessed a rapid development in synthesizing speech for single-language text due to the introduction of sequence-to-sequence models \cite{DBLP:conf/interspeech/WangSSWWJYXCBLA17, shen2018natural, DBLP:conf/iclr/PingPGAKNRM18, zhang2020unsupervised} and high-fidelity neural vocoders \cite{vanwavenet, oord2018parallel, 48585,NEURIPS2019_6804c9bc }. However, multi-lingual TTS remains a challenging task. The main challenge lies in disentangling language attributes from speaker identities in order to achieve code-switching and cross-lingual voice cloning.

% Usually, text-to-speech (TTS) assumes text comes from one language. However, in the multi-lingual community, code-switching is a common phenomenon. Code-switching refers to switching between different languages in text or speech. Although subtle differences between code-switching and code-mixing exist, we use code-switching as a term to denote mixed-lingual text or speech in this paper.

% Code-switched TTS is more challenging than mono-lingual TTS. One big challenge is the low-resource data problem. 

Usually, multi-lingual speech from the multi-lingual speaker is required to build a TTS system that can perform code-switching and cross-lingual voice cloning. For example, \cite{Traber99frommultilingual} builds a bilingual TTS system using bilingual speech data from a bilingual speaker. However, it is hard to find a speaker who is proficient in multiple languages and has smooth articulation across different languages.  

Thus, some studies have started to build cross-lingual TTS systems using mono-lingual TTS data. \cite{1415035} uses a mixture of mono-lingual speech data to build a Hidden-Markov-Model(HMM) based code-switched TTS system, where HMM states are shared across different languages.  \cite{4730269, 5153557} construct an HMM-based Mandarin-English TTS system with shared context-dependent HMM states, and mapping from the bilingual speech is learned. \cite{He2012TurningAM} turns a mono-lingual speaker into multi-lingual for mixed-lingual TTS by formant mapping based frequency warping, adapting F0 dynamics, and adjusting speaking rates accordingly. \cite{Sitaram2016ExperimentsWC} presents a code-mixed TTS system where mappings between the phonemes of both languages are used to synthesize the mixed-lingual text. \cite{Chandu2017Speech} develops a bi-lingual TTS system for navigation instructions using a mixture of mono-lingual speech datasets and a unified phone set for two languages.

With a successful application of sequence-to-sequence models in TTS \cite{DBLP:conf/interspeech/WangSSWWJYXCBLA17, shen2018natural, gibiansky2017deep, DBLP:conf/iclr/PingPGAKNRM18}, some researchers have begun to investigate sequence-to-sequence cross-lingual TTS.

\cite{8682674, 8682927, Nekvinda2020OneMM, DBLP:conf/interspeech/XueSXXW19, zhan2021improve, zhang2021revisiting, zhan2021exploring}, build a sequence-to-sequence code-switched TTS system using a mixture of mono-lingual speech data. \cite{8682674} proposes to use bytes as model inputs instead of grapheme, resulting in synthesizing fluent code-switched speech; but the voice switches for different languages. \cite{8682927} explores two kinds of encoders to handle alphabet inputs of different languages, namely (1) shared multi-lingual encoder with explicit language embedding; (2) separate mono-lingual encoders for each language. \cite{Nekvinda2020OneMM} introduces meta learning to improve multi-lingual TTS based on \cite{8682927}. \cite{DBLP:conf/interspeech/XueSXXW19} builds a mixed-lingual TTS system by pre-training an average voice model trained by multi-speaker mono-lingual data. \cite{DBLP:conf/interspeech/XueSXXW19} also looks into the effect of position of speaker embedding on speaker consistency and phoneme embedding on intelligibility and naturalness. \cite{zhan2021improve} use pitch contour to improve cross-lingual TTS performance. \cite{zhang2021revisiting, zhan2021exploring} investigates cross-lingual TTS using IPA symbols.

\cite{48331} proposes to build a multi-lingual TTS system using hundreds of hours of high-quality TTS data in three languages. Therefore, the system can perform code-switched TTS and cross-lingual voice cloning. However, \cite{48331} shows that it is difficult to achieve cross-lingual voice cloning if only one speaker is available for each language in training data, even when augmented with the proposed speaker-adversarial loss which aims to disentangle textual representation from speaker identities.

% Although the previous works above have promisingly achieved a improvement in code-switches TTS or cross-lingual voice cloning. However, it is both expensive and laborious to record a large amount of high-quality TTS data from many mono-lingual speakers as in \cite{DBLP:journals/corr/abs-1904-06063, zhang2019learning}. 

% These methods have promisingly achieved a better result in code-switched speech synthesis or cross-lingual voice cloning. However, training the model with only a mixture of mono-lingual speech in different languages still suffers from distribution mismatch, both textually and acoustically. As a result, during inference, the code-switched text and speech are unseen for the trained model.

% Although it is hard to record bilingual speech data for a bilingual speaker,

In this paper, we aim to achieve cross-lingual voice cloning using low-quality code-switched found data. As it is both laborious and expensive to record multi-lingual TTS data as in \cite{Traber99frommultilingual} or a large amount of high-quality mono-lingual TTS data as in \cite{DBLP:conf/interspeech/XueSXXW19, 48331}, we propose to utilize abundant code-switched found speech data to achieve cross-lingual voice cloning for the target speakers. The contributions of this work are summarised as below:

\begin{itemize}
\item To the best of our knowledge, it is the first work to improve cross-lingual voice cloning using low-quality code-switched found data.
\item Our proposed method to use low-quality code-switched found data can significantly improve the performance of cross-lingual voice cloning, achieving a comparable result to the SOTA performance.
\item Experiments show that our proposed method can also be incorporated with using separate encoders \cite{8682927} to improve the results, which indicates its great compatibility with other methods in cross-lingual TTS.
\end{itemize}

The rest of the paper is organized as follows:  Section 2 reviews some related works. Section 3 describes the baseline models and our proposed approach. Section 4 details the experimental setup and result analysis. The paper is closed with a conclusion in Section 5.

\section{Related Works}

There are some researches on building TTS models using low-quality found data. \cite{cooper2017utterance} investigates automatically selecting ASR speech data to improve the intelligibility of TTS systems. Similarly, \cite{kuo2018data, kuo2019selection, baljekar2016utterance} also find that careful data selection could improve the performance of TTS models using found data.

This work is related to the works above since this work investigates improving the TTS performance using low-quality found data. However, the previous works focus on mono-lingual TTS using mono-lingual found data. This work aims to improve cross-lingual voice cloning performance by utilizing code-switched found data. This task is more challenging because we not only need to deal with the ``low-quality" attribute of these found data, but also to handle the data mismatch problems described in section \ref{motivation}.

\begin{figure}[t]
  \centering
  \includegraphics[ width=\linewidth, ]{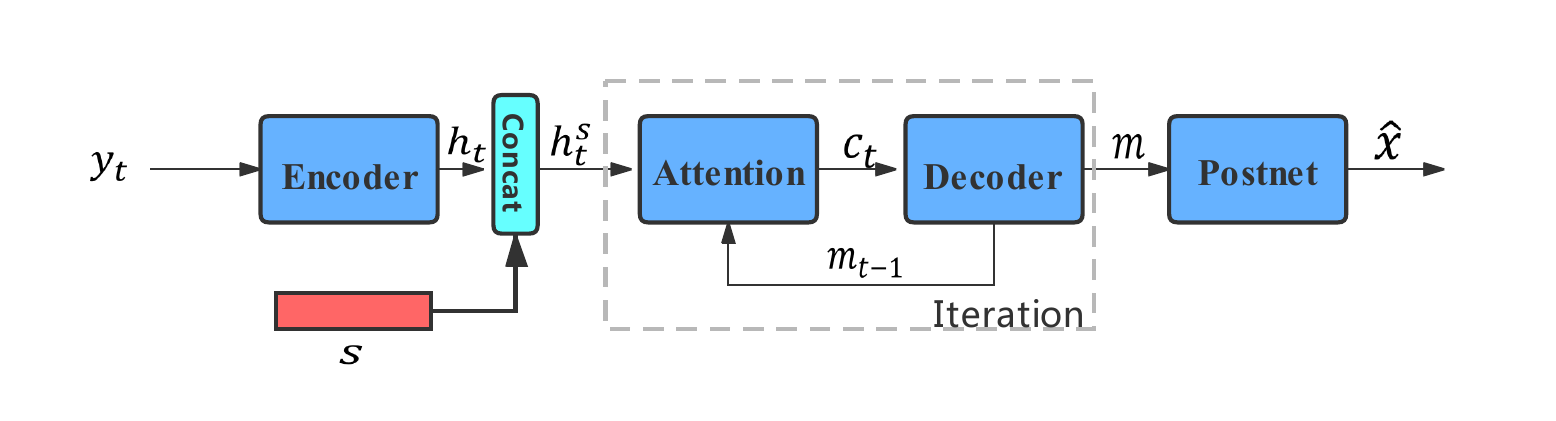}
  \caption{The model architecture studied. }
  \label{fig:model}
\end{figure}

\section{Methodology}

\subsection{The baseline model}

In this paper, we use Tacotron2 \cite{DBLP:conf/interspeech/WangSSWWJYXCBLA17} as the baseline TTS model structure, which is a SOTA sequence-to-sequence model to generate mel spectrogram given the text sequence. We make some modifications to the vanilla Tacotron. We replace the Location-Sensitive Attention (LSA) in the vanilla Tacotron with Gaussian Mixture Model (GMM) Attention for more robust sequence generation. We then augment the model with additional speaker embedding for modeling speaker characteristics. The whole model is illustrated as Figure \ref{fig:model} and below equations.

\begin{equation}
\begin{aligned}
        \centering
      h_t ={}&  Encoder(y_{t} )   \\
      h_t^s = {}& Concat (h_t, s) \\
      h^{att}_{t}, c_{t}, o_{t} = {} & Attention(m_{t-1},  h_t^s,h^{att}_{t-1}, c_{t-1} )\\
      h^{dec}_{t}, m_{t} ={}& Decoder( c_{t}, o_{t},h^{dec}_{t-1} )\\
      \hat{x} ={} & Postnet(m)
\end{aligned}
\end{equation}

where $h_t$ , $y_t$, and $s$ are the encoder output, phoneme input and speaker embedding, respectively. Concat denotes embedding concatenation. $h^{att}_{t}$ and $h^{dec}_{t}$ are hidden representation of attention RNN and decoder RNN. $c_{t}$, $o_{t}$ are context vector of attention and output of attention RNN. $m$ and $\hat{x}$ are mel spectrogram before and after postnet.

\subsection{Separated mono-lingual encoder}
We also implement separate mono-lingual encoder (SPE) \cite{8682927} to verify the compatibility of our proposed approach with other methods proposed for cross-lingual TTS. We briefly review SPE as below.

\begin{figure}[t]
  \centering
  \includegraphics[ height=7cm]{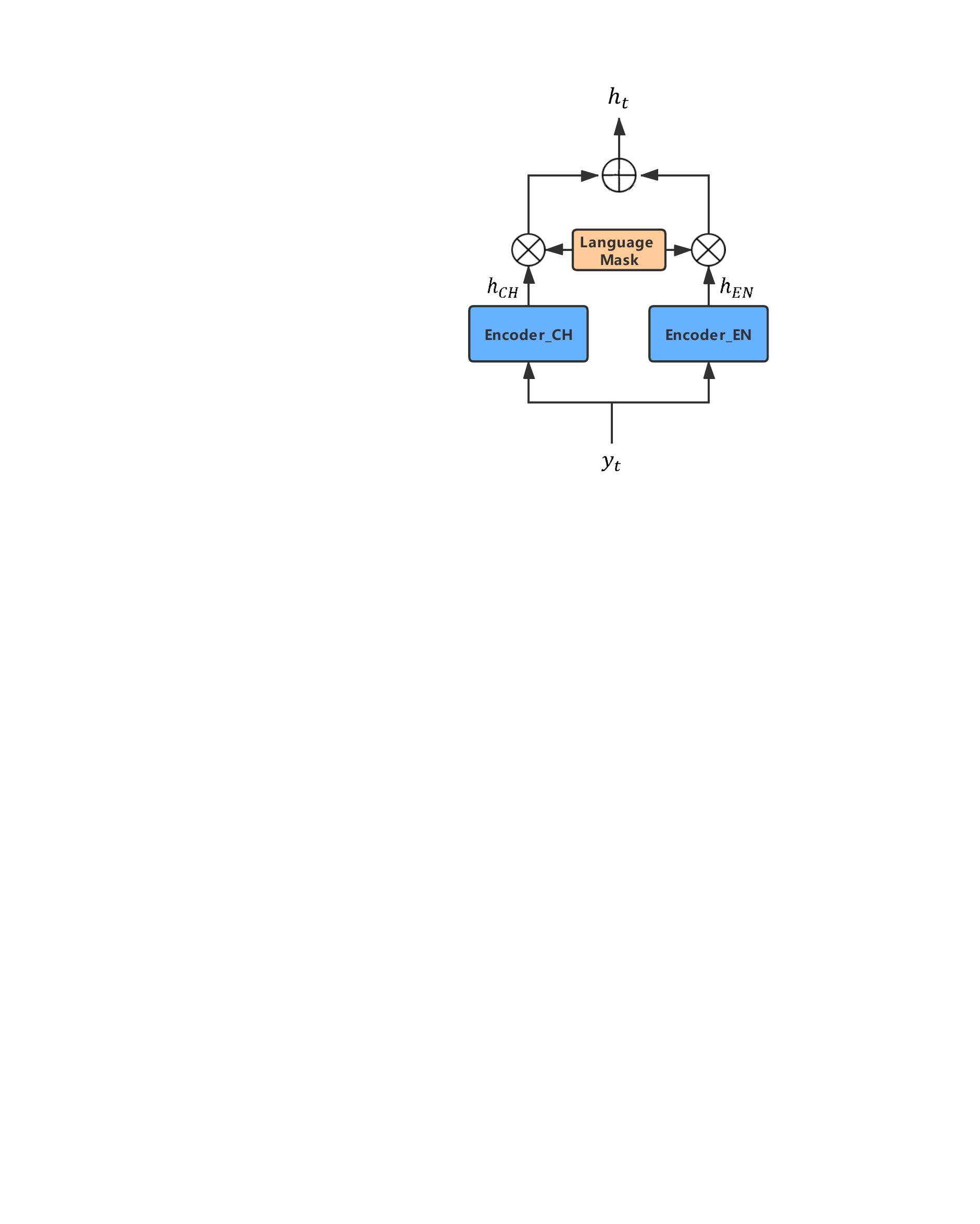}
  \caption{The encoder of SPE model proposed in \cite{8682927}. }
  \label{fig:spe}
\end{figure}

To avoid the mutual interference between the representations of different languages, the separate mono-lingual encoder (SPE) system \cite{8682927} uses two separate encoders for inputs from two languages. The whole structure of the encoder is illustrated in Figure \ref{fig:spe}. The model has an English encoder $Encoder_{\_EN}$ and a Chinese encoder $Encoder_{\_CH}$. The input character sequence is fed into both encoders, and a language ID sequence $L$ is used as a mask to extract the language portion from the respective encoder. The final encoder output is computed as follows :

\begin{equation}
% \centering
\begin{aligned}
\centering
        h_t  =  & Encoder_{\_CH}(y_t)   \otimes Mask_{CH} \oplus \\  
                &  Encoder_{\_EN}(y_t)   \otimes Mask_{EN}
\end{aligned}
\end{equation}

where $\otimes$ refers to element-wise multiplication, and $\oplus$ denotes element-wise addition. During training, each encoder block is actually trained by the corresponding mono-lingual data, since either $Mask_{CH}$ or $Mask_{EN}$ is set into 1. But in the testing phase, as stated in \cite{8682927}, it is better to input the whole input sequence into both encoder blocks for computing the textual presentation. Although the encoder would mishandle the inputs from the other language, maintaining the whole context information could help reduce the mismatches in the language boundaries and learn better alignment for code-switched utterances. The rest of the SPE model is implemented as the baseline Tacotron2 model.

% \subsubsection{Speaker adversarial training}

% \cite{zhang2019learning} proposed to employ a speaker adversarial layer to encourage the encoder to learn speaker- and language-independent textual representation. To achieve that, \cite{zhang2019learning} imposed a speaker classifier on the encoder output and inserted a gradient reversal layer prior to the classifier. 

\subsection{Proposed Method}

\subsubsection{Motivation} \label{motivation}
There is a data distribution mismatch problem in performing cross-lingual voice cloning using mono-lingual data. Namely, we only have mono-lingual data from target speakers when training the model, while we want to synthesize multi-lingual speech using the voices of target speakers when testing. In this paper, we hypothesize that using code-switched data could mitigate this mismatch problem. Since it is hard to record code-switched data, we choose to collect low-quality code-switched data in this paper.

Although it is more convenient to collect low-quality mono-lingual speech data than code-switched speech data, \cite{48331} has found that even training the model using hundreds of high-quality mono-lingual data can not achieve satisfactory performance on speaker similarity in cross-lingual voice cloning, especially when Chinese speakers are involved. That is to say, simply using abundant mono-lingual data could not disentangle language attributes from speakers well. Thus, in this paper, we hypothesize that utilizing code-switched speech data could help better disentangle language attributes from speakers than mono-lingual data.

\subsubsection{Training \& filtering strategy} \label{train}

Due to the ``low-quality" attribute of these code-switched found data, we adopt a pretrain-finetune two-stage training strategy. Inspired by \cite{DBLP:conf/interspeech/XueSXXW19}, we pretrain the TTS model using a mixture of code-switched found data and high-quality TTS data from target speakers. When pretraining, we assign each speaker a speaker embedding for multi-speaker training, as we find this pretraining method can achieve better results than treating all the data as from a single speaker and using no speaker embedding in our initial experiment.
When fine-tuning the model, we fix the encoder and only fine-tune the rest of the model, since we find that fine-tuning the whole model would cause the problem of catastrophic forgetting. \footnote{Forgetting about disentangling the language attributes from speaker identities, since the fine-tuning data are mono-lingual.}  

Moreover, modeling seq-to-seq TTS models using low-quality speech data would be challenging, since the attention mechanism would fail to learn a robust alignment between text representations and acoustic features. Thus, inspired by \cite{cooper2016data,cooper2017utterance}, we filter some potentially ``noisy'' data by metrics including fast speaking rate, hypo-articulation, and signal to noise ratio (SNR).

\begin{table}
    \caption{Word error rate (WER) of synthesized utterances.}
\centering
\begin{tabular}{ c c c c c }
\hline\hline

SPK &  Model &  CS-Target  & CN-Target & EN-Target \\
\hline
 &  GT  &  12.64\%  &   0\%   &  4.48\%   \\
\hline
\multirow{4}{*}{CN} 
& Tac      
& 59.20\% 
& \textbf{0.1\%} 
& Fail   \\

& Tac\_Mix 
& \textbf{15.52\%}  
& 0.7\% 
& \textbf{18.97\%}   \\

\cline{2-5} 

& SPE 
& 27.27\%    
& 3.07\%   
& 33.05\% \\

& SPE\_Mix 
& \textbf{16.82\%}     
& \textbf{1.23\%}  
& \textbf{23.73\%}  \\

% \cline{2-5}
\hline

\multirow{4}{*}{EN} 
& Tac 
& 54.60\%  
& Fail 
& \textbf{5.17\%}  \\

& Tac\_Mix 
& \textbf{28.83\%} 
& \textbf{22.61\%} 
& 6.21\%   \\

\cline{2-5} 

& SPE 
& 30.68\%      
& 59.42\%
& 6.78\% \\

& SPE\_Mix 
& \textbf{29.55\%}      
& \textbf{22.90\%}   
& \textbf{6.78\%} \\

\hline
\end{tabular}
\label{table:WER}
\end{table}

\begin{table*}
    \caption{Naturalness and speaker similarity MOS of synthesizing CS, CN, and EN text using CN-SPK and EN-SPK. We also report the cross-lingual cloning results in \cite{48331} }
\centering
\begin{tabular}{ c c c c c c c c }
\hline\hline

\multirow{2}{*}{ Source Speaker} &  \multirow{2}{*}{Model} & \multicolumn{2}{ c }{CS-Target} & \multicolumn{2}{ c }{CN-Target}  & \multicolumn{2}{ c }{EN-Target}\\
\cline{3-8}

&  & Naturalness & Similarity  & Naturalness & Similarity & Naturalness & Similarity  \\
\hline

&  Ground-Truth  
&  $ 4.20 \pm 0.14 $  
&  $ 4.21 \pm 0.09 $  
&  $ 4.49 \pm 0.19 $  
&  $4.36  \pm 0.22 $  
&  $4.73  \pm 0.12 $ 
&  $4.28  \pm 0.08 $  \\
\hline

\multirow{5}{*}{CN-SPK} 

& Tac 

&  $ 3.11  \pm 0.11 $   
&  $ 3.49  \pm 0.14 $  
&  \textbf{ 4.23 $ \pm $ 0.08 }   
&  $ 4.01  \pm 0.08 $   
&  Fail               
&  Fail \\

&  Tac\_Mix 
&  \textbf{ 4.30 $\pm$ 0.07 }  
&  \textbf{ 4.18 $\pm$ 0.08 } 
&  $ 4.15 \pm 0.08 $  
&  \textbf{ 4.26 $\pm$ 0.07 } 
&  \textbf{ 3.73 $\pm$ 0.09 }  
&  \textbf{ 3.83 $\pm$ 0.08 } \\

\cline{2-8}

& SPE 
&  $ 3.40 \pm 0.08 $  
&  $ 3.45 \pm 0.10 $  
&  $ 4.06 \pm 0.09 $  
&  $ 4.15 \pm 0.10 $ 
&  $ 2.63 \pm 0.13 $  
&  $ 2.93 \pm 0.13 $ \\

& SPE\_Mix 
&  \textbf{ 4.02 $\pm$ 0.08 }  
&  \textbf{ 3.98 $\pm$ 0.06 }  
&  \textbf{ 4.14 $\pm$ 0.13 }  
&  \textbf{ 4.15 $\pm$ 0.08 } 
&  \textbf{ 3.29 $\pm$ 0.13 }  
&  \textbf{ 3.83 $\pm$ 0.09 } \\

\cline{2-8}

& Google's \cite{48331} 
& -  
& - 
& - 
& - 
& $4.49 \pm 0.08 $ 
& $2.46 \pm 0.10 $\\

% \cline{2-8}
\hline

\multirow{5}{*}{EN-SPK} 
& Tac 
&  $2.95  \pm 0.07 $  
&  $ 2.28 \pm 0.09 $  
&  Fail              
&  Fail  
&  \textbf{4.41  $\pm$ 0.07 }  
&  $ 3.90 \pm 0.08 $ \\

&  Tac\_Mix 
&  \textbf{ 4.05 $\pm$ 0.07}   
&  \textbf{ 4.01 $\pm$ 0.05}  
&  \textbf{ 3.83 $\pm$ 0.08}   
&  \textbf{ 3.95 $\pm$ 0.08}  
&  $ 4.39 \pm 0.05 $  
&  \textbf{ 4.08 $\pm$ 0.06 } \\

\cline{2-8}

& SPE 
&  $ 3.53 \pm 0.07 $  
&  $ 2.95 \pm 0.07 $  
&  $ 2.73 \pm 0.10 $  
&  $ 3.40 \pm 0.10 $ 
&  \textbf{ 4.33 $\pm$ 0.08 }  
&  $ 3.91 \pm 0.07 $ \\

& SPE\_Mix 
&  \textbf{ 3.92 $\pm$ 0.05 } 
&  \textbf{ 3.82 $\pm$ 0.10 }  
&  \textbf{ 3.77 $\pm$ 0.07 }  
&  \textbf{ 3.92 $\pm$ 0.08 } 
&  $ 4.27 \pm 0.08 $  
&  \textbf{ 3.93 $\pm$ 0.07 } \\

\cline{2-8}

& Google's \cite{48331}
& -  
& - 
& $3.94 \pm 0.09 $ 
& $3.03 \pm 0.10 $  
& - 
& - \\

\hline
\end{tabular}
\label{table:MOS}
\end{table*}

\section{Experiments}
\subsection{Experimental setup}
\subsubsection{Data}
In this paper, we aim to utilize low-quality code-switched found data to enable our mono-lingual target speakers to speak foreign languages (a.k.a. cross-lingual voice cloning) . We use only mono-lingual speech data from two female speakers. The first one is a Chinese female speaker from our internal corpus, and the other is the English female speaker from \cite{ljspeech17}. The number of training utterances for each speaker is 5000, with in total 10 hours of speech. For the low-quality code-switched data, we use the code-switched corpus of the ASRU 2019 Code-Switching Challenge. Since the corpus is designed for automatic speech recognition, the quality is significantly lower than TTS data. As described in section \ref{train}, we perform filtering on the original corpus to select only 100 speakers for experiments. For each speaker, we further filter 10\% lowest-quality data. As a result, we used in total 33000 code-switched utterances (about 27-hour data). 200 utterances of each type (i.e Chinese, English and Code-switched) are randomly selected for evaluations, which are not used in training and development.

\subsubsection{Training Setup}

In this paper, we build several TTS systems as follows:

\begin{itemize}
    \item Tac : Tacotron2 model trained with mono-lingual data;
    \item Tac\_{Mix} : Tacotron2 pre-trained with mixture of data, then fine-tuned with mono-lingual data;
    \item SPE : SPE TTS system trained with mono-lingual data;
    \item SPE\_{Mix} : SPE TTS system pre-trained with mixture of data,  then fine-tuned with  mono-lingual data;
    % \item SAT\_Mix : Tacotron-based TTS system augmented with speaker adversarial training 
    % pre-trained with mixture of data, then fine-tuned with the mono-lingual data;
    % \item SPE\_SAT\_\_Mix : Tacotron-based SPE TTS system augmented with speaker adversarial training 
    % pre-trained with mixture of data, then fine-tuned with the mono-lingual data.
\end{itemize}

% To provide comparisons to the SOTA performance in cross-lingual voice cloning, we include the results in \cite{zhang2019learning} as a reference \footnote{ These results can be seen the state-of-the-art (STOA) performance in cross-lingual voice cloning. But, since they used the internal data in experiments, we can not do fair comparison under the same experimental setting. Thus, we only report the results stated in the paper as a reference.}. 

For Model Tac and SPE, we train the models for 200k steps. For Model Tac\_{Mix} and SPE\_{Mix}, we pre-train the models for 100k steps. Then the models are fine-tuned using data from target speakers. Early stopping is used to avoid over-fitting. We train the models using the Adam \cite{DBLP:journals/corr/KingmaB14} optimizer and a batch size of 32. We use an initial learning rate of 1e-3, which is halved to 4 $×$ 1e-4 until convergence.

% We use 
% gradient clipping with factor 0.5 to the gradient reversal layer as in \cite{zhang2019learning}.

Waveforms are synthesized by WaveNet vocoder which generates 16-bit speech at a 16kHz sample rate conditioned on the predicted spectrograms. We used a single variance-bounded Gaussian distribution to model the waveform samples as in \cite{ping2018clarinet}, which could relieve the quantization errors in synthetic speeches brought by previous categorical distribution. We train one vocoder for each target speaker. For training the vocoder for the Chinese speaker, we use 10-hour speech data from the target speaker, while for the English one, we use 15-hour data from the target speaker.

% \subsection{Research questions}

\subsection{Result Analysis}

% \footnote{We used WER for English utterances, CER for Chinese utterances, and a mixture of WER and CER for Chinese-English code-switch utterances.} 

To verify whether using low-quality code-switched data is helpful, both objective and subjective tests are carried out. For the objective test, we use iFLYTEK speech-to-text API \footnote{https://www.xfyun.cn/} to recognize the generated speech and use the word error rate (WER) as the measurement metric for intelligibility. For the subjective test, we conduct a formal listening test. In the listening test, 16 raters are included, who are native Chinese Mandarin speakers and proficient in English. The listeners are asked to evaluate the naturalness and speaker similarity of synthesized speech by each model using the mean opinion score (MOS). Scores range from 1 to 5 with an interval of 0.5. We involve three types of utterances in the listening test. Each type of utterances includes 10 samples, (namely, 10 Mandarin-English code-switched utterances (CS-Target), 10 Chinese utterances (CN-Target), and 10 English utterances (EN-Target).). Each utterance is rated by all 16 listeners for naturalness and similarity. The results of the ground-truth speech are also provided as references. Speech demos are available at  \url{https://haitongzhang.github.io/Code-switch-TTS/} .

\subsubsection{Objective evaluation}

For the objective evaluation, we use about 5-minute speech data for each type. The results are provided in Table \ref{table:WER}. 

It is clearly shown that using low-quality code-switched data almost does not hurt synthesizing source language (namely, synthesizing Chinese speech using the Chinese speaker's voice and English speech using the English speaker's voice). 

We find that utilizing these low-quality data to pretrain the model could significantly reduce the WER in case of code-switching, although the improvement on the Tacotron model is more significant than on SPE. 

In cases of cross-lingual voice cloning, since model Tac fails to perform cross-lingual voice cloning, we do not compute the WER of the model. However, we find that model SPE\_Mix outperforms significantly SPE system, which indicates the effectiveness of using low-quality code-switched data. 

%Meanwhile, it is found that the reduction on the English speaker is smaller than that on the Chinese speaker. We speculate that since the low-quality code-switched speech data are all from Chinese speakers, the improvement on the Chinese speaker would be more significant. 

Moreover, we find that model Tac\_Mix slightly outperforms SPE\_Mix because model Tac\_Mix uses phonemes as inputs while SPE\_Mix uses characters as inputs. As a result, when training, SPE\_Mix needs to learn the actual pronunciations of characters simultaneously, which is more difficult.

\subsubsection{Subjective evaluation}

The results of subjective evaluation on naturalness and speaker similarity are provided in Table \ref{table:MOS}. 

For synthesizing source language speech, pre-training using low-quality code-switched data almost does not have a negative impact on the results, which is consistent with the results of objective evaluation.

In the case of code-switched speech, Model Tac can hardly generate intelligible code-switched speech regardless of the speaker, since the naturalness MOS scores are only about $3.0$. With pre-training using low-quality code-switched data, Model Tac\_Mix can synthesize natural code-switched speech. Specifically, compared with Model Tac, Model Tac\_Mix provides an increase of $38\%$ and $37\%$ in the naturalness MOS score for the Chinese speaker and the English speaker, respectively. Meanwhile, the improvement in speaker similarity is significant, with an increase of $20\%$ and $76\%$, respectively.

As shown in Table \ref{table:MOS}, using separate encoders for two different languages can improve the performances in code-switched synthesis, which is consistent with \cite{8682927}. When further pre-training the model using low-quality code-switched data, Model SPE\_Mix outperforms Model SPE, with an increase of $18\%$ and $11\%$ in naturalness and $15\%$ and $30\%$ in speaker similarity, respectively.

As far as cross-lingual voice cloning is concerned, Model Tac fails to perform cross-lingual voice cloning. We find that when with the Chinese speaker embedding, the generated English utterances still sound like the English speaker and vice versa. But Model Tac\_Mix can achieve promising performance in cross-lingual voice cloning, which reflects the effectiveness of pre-training using low-quality code-switched data.

Besides that, using separate encoders, Model SPE can hardly perform cross-lingual voice cloning, since the performance is not satisfactory enough. But, when pre-training the model using low-quality code-switched data, Model SPE\_Mix can achieve a significantly better result in both naturalness and similarity, which indicates the proposed pretraining method using low-quality code-switched can be incorporated with SPE system to achieve better results.

% What's more, we found that Model SPE\_Mix gains no advantage over Model Tac\_Mix, which indicates the separated encoders are not necessary if pre-training is adopted. By comparing the performances between Model Tac\_Mix and Model SPE, we concluded that pre-training the model with code-switched data is a more effective approach that using separated encoders for different languages. 

Moreover, we report the cross-lingual cloning results in \cite{48331}. Although the naturalness of our proposed system is not better than that of \cite{48331} \footnote{Partially because of the quality of our training data}, the speaker similarity of our proposed best system is significantly better than that of \cite{48331}. Generally speaking, our system has achieved performance comparable to the SOTA system considering the differences in experimental settings (e.g., the quality and quantity of training data).

% \subsubsection{The effect on types of code-switch text}

% We also investigate whether the ability can be fully transferred in terms of types of code-switch text. Specially, we group code-switch text into two groups. The first group refers to the code-switch text sequence with English words embedded in the middle or at the end of the text sequence, while the second group represents those which starts with English words. %The sample texts are given here:

% \begin{CJK*}{GBK}{song}

% \begin{enumerate}
    % Group1 Example: 他的研究领域是 theoretical science
    % Group2 Example: Hamburger menu 的替代方案
% \end{enumerate}
% \end{CJK*}

% The results of MOS for two types of code-switch text are provided in Figure \ref{fig:mos-type-spk2} and Figure \ref{fig:mos-type-spk1}. For the Chinese speaker (Figure \ref{fig:mos-type-spk2}), without transfer learning, the results of two types of code-switch text are close. With pre-training, the naturalness of the synthesis of type0 text is higher than that of type1 text. We attributed this to the imbalanced distribution of these two-type data in the pre-training code-switch dataset. We found a similar case for the English speaker, except for Model Pre1. The reason is that semi-supervised pre-training almost fails to transfer the ability to the English speaker.

\section{Conclusion}
In this paper, we aim to achieve cross-lingual voice cloning using mono-lingual data from target speakers. We leverage abundant low-quality code-switched found data to pretrain the TTS model. We conclude our findings as followed:

\begin{itemize}
    \item When utilizing low-quality code-switched found data, data filtering strategy and pretrain-finetune training strategy are helpful to mitigate the attribute of ``low-quality", as it does not have a negative impact on synthesizing source speech.
    \item Pretrained with low-quality  code-switched data, the model can improve the performance in case of code-switched synthesis.
    
    \item Pretrained with low-quality code-switched data, the model can achieve comparable performance to the SOTA model in cross-lingual voice cloning.
    
    \item Pretaining with low-quality code-switched data can be incorporated with using separate encoders to bring about a further improvement.
\end{itemize}

Although experiments have shown that pre-training with low-quality code-switched found data is useful to achieve cross-lingual voice cloning, we only investigate using paired code-switched data in this paper. However, there are more unlabelled code-switched data in the wild. Future investigation should be carried out on unsupervised pre-training using these unlabelled data.
% by utilizing the code-switched data from ASR. The experiments indicate that with transfer learning, the model can synthesize natural code-switched speech with a high speaker consistency. We also found that the improvement brought by transfer learning is more significant that that of the separated encoders. Another interesting finding is that if pre-training is adopted, using separated encoders for different language can not bring further improvement. 

% In the case of cross-lingual cloning, we concluded that with pre-training the model with code-switched data from ASR, the system can achieve the performance comparable to the SoTA system.

% Although a significant improvement is achieved by pre-training the model, such paired pre-training code-switched data may not exist in the wild or for other spoken languages. Thus, for the future works, we plan to investigate using unpaired code-switched data for pre-training and involve more languages in the experiments.

% \section{ACKNOWLEDGMENTS}

\bibliographystyle{IEEEtran}

\bibliography{refs}

\end{document}